\documentstyle[prl,twocolumn,aps]{revtex}
\input{epsf}
\begin{document}
\draft
\twocolumn[\hsize\textwidth\columnwidth\hsize\csname @twocolumnfalse\endcsname
\title{Phase of bi-particle localized states for the 
Cooper problem in two-dimensional disordered systems}

\author{J. Lages and D. L. Shepelyansky$^{*}$}

\address {Laboratoire de Physique Quantique, UMR 5626 du CNRS, 
Universit\'e Paul Sabatier, F-31062 Toulouse Cedex 4, France}

\date{April 11, 2001}

\maketitle

\begin{abstract}
The Cooper problem  is studied numerically 
for the Anderson model with disorder in two-dimensions.
It is shown that the attractive Hubbard interaction 
creates a phase of bi-particle localized states 
in the regime where non-interacting states are
delocalized. This phase cannot be obtained in the mean-field
approximation and the pair coupling energy is 
strongly enhanced in this regime. The effects of magnetic
field are studied and it is shown that under certain
conditions they lead to delocalization. 

\end{abstract}
\vskip1pc]

\narrowtext

\section{Introduction}
Recently a great deal of attention has been attracted to investigation
of superconductor-insulator transition (SIT) in systems with
disorder. Various approaches are used to study this problem including
analytical theoretical methods \cite{larkin,matveev}, intensive 
numerical simulations of many-body quantum systems 
with quantum Monte Carlo methods \cite{erik94,trivedi1}, as well
as mean field numerical simulations \cite{franz,trivedi2,trivedi3}. These
theoretical studies are stimulated by challenging experiments on
SIT in disordered films \cite{goldman,gantmakher} and high-$T_c$ 
superconductors \cite{boebinger,ando}. The results obtained in 
\cite{boebinger} show an interesting correlation between the optimal
doping and the Anderson transition in the normal phase obtained
by application of a strong pulsed magnetic field. Even if the experiments 
\cite{boebinger,ando} are done with three-dimensional crystals
the coupling between two-dimensional planes is relatively weak
and the two-dimensional effects should play an important role.
Due to that it is relevant to study the SIT in two-dimensional
disordered systems. In the case of weak disorder the Anderson 
theorem \cite{anderson,abrikosov} guarantees that the 
superconductivity is not affected by disorder. However it is 
not obvious if the theorem is still valid in the presence
of relatively strong disorder. It is quite possible that in 
this regime the interplay of disorder and interaction can lead
to appearance of new physical effects. The theoretical 
investigation of this regime is however rather difficult. 
The existent analytical methods are not well adapted to
the regime of strong interaction and disorder. At the same
time the numerical studies also meet serious difficulties.
Indeed the direct diagonalization methods are restricted to 
relatively small system size since the Hilbert space grows
exponentially with the number of particles 
\cite{poilblanc,dagota}. The quantum Monte Carlo methods 
are not so sensitive to a huge size of the Hilbert space but
still they are restricted to systems of quite moderate size
(for example lattices of 8x8 sites in \cite{trivedi1}).

In the view of above numerical difficulties it is natural
to develop the approach introduced by Cooper \cite{cooper}
and to study the problem of two particles with attractive
interaction near the frozen Fermi sea in the presence of
disorder. Even if the original Cooper problem of two particles 
without disorder does
not reproduce exactly the BCS theory of many-body
problem it nevertheless captures the essential physical 
properties of the system and gives the appearance of coupled
states with a qualitatively correct coupling energy and
correlation length. Without disorder the Cooper problem can
be solved exactly. However in presence of disorder the 
situation becomes more complicated even for only two 
interacting particles (TIP). Indeed, for relatively
strong disorder even the matrix elements of interaction
between non-interacting eigenstates can not be obtained
analytically and due to that the problem should be studied
numerically. The first numerical studies of the Cooper
problem in the presence of disorder were done in
\cite{lages} for two particles with attractive Hubbard
interaction in the three-dimensional Anderson model.
These studies showed that the interaction can lead to
localization of pairs in the non-interacting metallic phase.
This result is qualitatively different from the mean field
solution of the Cooper problem in the presence of disorder
(Cooper ansatz) which gives delocalized pairs for the
same parameters. This shows that the non-diagonal 
interaction induced matrix elements play an important role
and lead to new physical effects which are not captured
by the mean field approximation. 

In this paper we study the Cooper problem on a two-dimensional 
lattice with disorder described by the Anderson model. Our
numerical studies show that near the Fermi level
the attractive Hubbard interaction
between two particles creates localized pairs in the regime
where non-interacting eigenstates are well delocalized (extended).
The coupling energy of these pairs is much larger than the 
coupling energy given by the mean field solution (Cooper ansatz).
Therefore energetically it is more favorable to have an insulator
with localized pairs instead of usual  weakly coupled 
delocalized Cooper pairs. This result indicates the appearance
of a new phase of bi-particle localized states (BLS phase) which
appears in the regime when non-interacting states are extended
(metallic). It is in a qualitative agreement with the quantum Monte Carlo
studies obtained recently in Toulouse \cite{bhargavi}.
This BLS phase is qualitatively different from
BCS solution which corresponds to weakly coupled delocalized pairs.
The paper is organized as follows.
The properties of BLS phase without magnetic field are discussed 
in Section 2. The effects of perpendicular magnetic field on
the ground state properties in the presence of interaction
and disorder are analyzed in Section 3. The discussion of
the results is presented in the last Section.

\section{Ground state properties without magnetic field}

To study the Cooper problem of two in\-terac\-ting particles in
the presence of disorder we use the two-dimensional Anderson
model. In this model the one-particle eigenstates are determined
by the Hamiltonian
\begin{equation}\label{HA}
H_{1}=\sum_{\mathbf n} E_{\mathbf n} \left\arrowvert
{\mathbf n}\left\rangle\right\langle{\mathbf n}\right\arrowvert
+ V\sum_{\langle\mathbf n,m\rangle} 
\left\arrowvert {\mathbf n}
\left\rangle\right\langle{\mathbf m}\right\arrowvert 
\end{equation}
where $\mathbf n$ and $\mathbf m$ are index vectors on the 
two-dimensional square lattice
with periodic boundary conditions, $V$ is the nearest neighbor hopping
term and the random on-site energies $E_{\mathbf n}$ are homogeneously
distributed in the energy interval 
$\left[-\frac{W}{2},\frac{W}{2}\right]$, where $W$ is the disorder
strength. This one-particle model has been extensively studied
by different authors, see for example \cite{mirlin}. For the two-particle
problem on this two-dimensional lattice we consider on-site attractive
Hubbard interaction between particles of strength $U<0$. We consider
the particles in the singlet state with zero total spin so that the
spatial wavefunction is symmetric with respect to particle permutation
(interaction is absent in the triplet state). 

To investigate the effects of interaction between particles near the
Fermi level we generalize the Cooper approach for the case with 
disorder. To do that we rewrite the TIP Hamiltonian in the basis
of one-particle eigenstates of the Hamiltonian (\ref{HA}). In this
basis the Schr\"odinger equation for TIP reads  
\begin{eqnarray}
\label{TIPSCH}
(E_{m_1}+E_{m_2})\chi_{m_1, m_2} & + & 
U \sum_{{m^{'}_1}, {m^{'}_2}} Q_{m_1, m_2, {m^{'}_1}, {m^{'}_2}}
 \chi_{{m^{'}_1}, {m^{'}_2}} \nonumber \\
  & = & E\chi_{m_{1}, m_{2}}.
\end{eqnarray}
Here $E_{m}$ are the one-particle eigenenergies corresponding to the 
one-particle eigenstates $\arrowvert\phi_m\rangle$ and 
$\chi_{m_1,m_2}$ are the components of the TIP eigenstate in the
non-interacting eigenbasis $\arrowvert\phi_{m_1}, \phi_{m_2}\rangle$.
The matrix elements $UQ_{m_1, m_2, {m^{'}_1}, {m^{'}_2}}$ gives
the interaction induced transitions between non-interactive 
eigenstates $\arrowvert\phi_{m_1}, \phi_{m_2}\rangle$ and 
$\arrowvert \phi_{m^{'}_1}, \phi_{m^{'}_2}\rangle$. These matrix
elements are obtained by rewriting the Hubbard interaction 
in the non-interactive 
\begin{figure}
\epsfxsize=8cm
\epsfysize=13cm
\epsffile{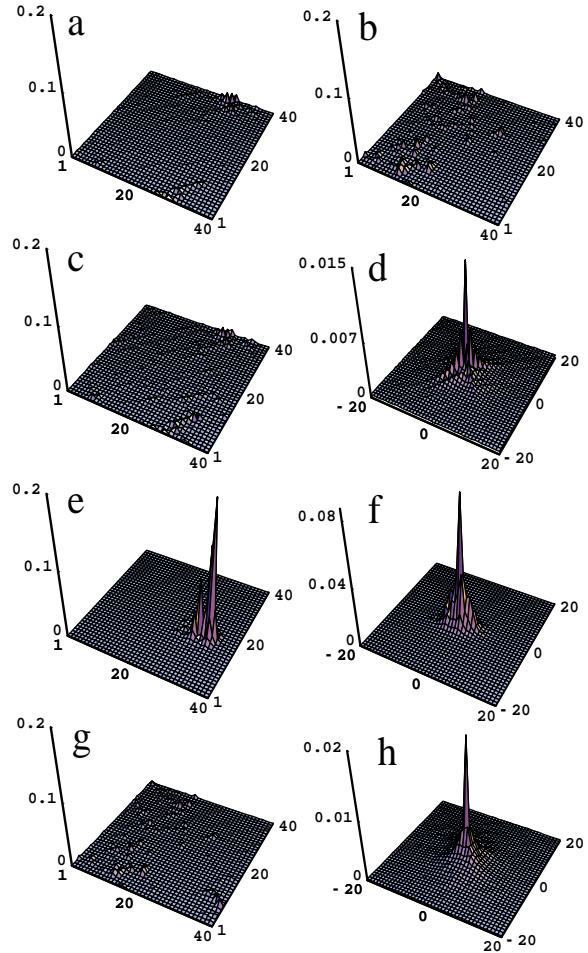}
\vglue 0.2cm
\caption{Ground state probability distributions
of two interacting particles for the Cooper problem with disorder
on a lattice of linear size $L=40$. The cases (a,b)
show one particle probability distribution $f(\mathbf{n})$
in absence of interaction $(U=0V)$
for the disorder strength $W=2V$ (a) and $W=5V$ (b).
All other cases are obtained for the Hubbard interaction $U=-2V$.
The cases (c,d) show the one-particle probability $f({\mathbf n})$
(c) and the interparticle distance probability 
$f_d(\mathbf{r})$ (d) for $W=2V$.
The same probabilities are shown in cases (e,f) for $W=5V$
($f({\mathbf n})$ for (e) and $f_d(\mathbf{r})$ for (f)).
The cases (g,h) present the probabilities for the same 
$W=5V$ as in cases (e,f) however here the ground state is obtained
in the mean field approximation of the Cooper ansatz.
All data are given for the same realization of disorder.}
\label{fig1}
\end{figure}
\noindent eigenbasis of model (\ref{HA}).
In the analogy with the original Cooper problem \cite{cooper}
the summation in (\ref{TIPSCH}) is done over the states above
the Fermi level with eigenenergies $E_{m^{'}_{1,2}}>E_F$ with 
$m^{'}_{1,2}>0$. The Fermi energy $E_F\approx 0$ is determined 
by a fixed filling factor $\nu =1/2$. To keep the similarity
with the Cooper problem we restrict the summation on $m^{'}_{1,2}$
by the condition $1< m^{'}_1+m^{'}_2\leq M$. In this way the 
cut-off with $M$ unperturbed orbitals introduces an effective
phonon frequency $\omega_D\propto M/L^2=1/\alpha$ where $L$
is the linear system size. When varying $L$ we keep $\alpha$
fixed so that the phonon frequency is independent of system
size. All the data in this work are obtained with $\alpha=15$
but we also checked that the results are not sensitive to
the change of $\alpha$. We note that a similar TIP model
was considered for the problem of two repulsive quasiparticles
near the Fermi level in \cite{imry,jacquod}. However, there
the studies were mainly addressed to the properties of excited
states while here we will investigate only the properties of
the ground state in the case of an attractive interaction.
We also note that the attractive case in one-dimension was
discussed in \cite{oppen,1dtip}.

To determine the characteristics of the ground state of the
generalized Cooper problem (\ref{TIPSCH}) we solved numerically 
the Schr\"odinger equation. After that we rewrite the obtained 
ground state
in the original lattice basis with the help of the relation
between lattice basis and one-particle eigenstates
$\arrowvert {\mathbf n}\rangle= \sum_m R_{{\mathbf n},m} 
\arrowvert \phi_m\rangle$. As a result of this procedure
we obtain the two-particle probability distribution 
$F({\mathbf n_1,n_2})$ from which we extract the one-particle 
probability 
$f({\mathbf n})=\sum_{\mathbf n_2}F({\mathbf n_1,n_2})$ 
and the probability of interparticle distance 
$f_d({\mathbf r})=\sum_{\mathbf n_2}F({\mathbf r+n_2,n_2})$
with ${\mathbf r=n_1-n_2}$. 

Typical examples of such probability distributions are 
presented in Fig. 1. Without interaction, at given disorder 
strength $W=2V$ and
$W=5V$ both particles are delocalized on the lattice of given
size $L=40$. In presence of interaction $U=-2V$ the ground state
remains delocalized for $W=2V$ and the probability distribution is
rather similar to the case of $U=0$ (compare Figs.1a and 1c).
On the contrary for $W=5V$ interaction completely changes 
the ground state properties leading to a clear localization
of both particles near each other (compare Figs.1b and 1e).
The wavefunction is localized in a rather compact way and the finite 
size of the lattice definitely does not affect this localization.
Fig.1f shows that in this localized state the particles remain
correlated being close to each other. This bi-particle localized 
ground state is obtained by exact diagonalization of (\ref{TIPSCH})
where all non-diagonal interaction induced matrix elements are
taken into account. It is interesting to compare this solution
with the mean field approximation (Cooper ansatz) in which only
diagonal terms are taken into account. Within the Cooper ansatz the particles
occupy the same non-interacting orbitals and only matrix elements
$Q_{m_1,m_2,m^{'}_1,m^{'}_2}$ with $m_1=m_2$ and $m^{'}_1=m^{'}_2$
are kept in (\ref{TIPSCH}). The ground state obtained from the 
Cooper ansatz is shown in Figs.1g and 1h and is clearly delocalized
contrarily to the strongly localized ground state  
obtained from exact diagonalization of (\ref{TIPSCH}) 
and shown in Figs.1e and 1f. 
In fact the ground state from the Cooper ansatz is more close to 
delocalized non-interacting eigenstate in Fig.1b than to the real
eigenstate in Fig.1e in the presence of interaction. The results
of Fig.1 definitely show that the attractive interaction leads
to localization of pairs in the regime when non-interacting states 
are delocalized. This localization is not captured by the Cooper
ansatz which neglects non-diagonal matrix elements and due to
that misses the essential physical effect. 

\begin{figure}
\epsfxsize=8cm
\epsfysize=8cm
\epsffile{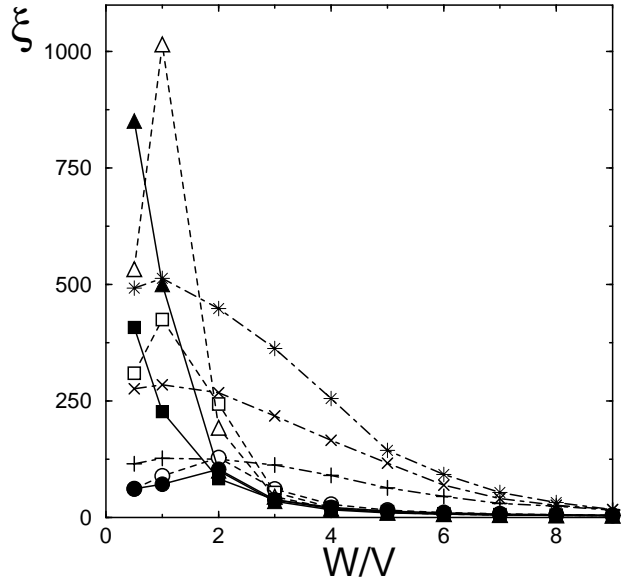}
\caption{Dependence of the inverse participation ratio 
$\xi$ of the TIP ground state on the disorder $W/V$.
The full lines with full symbols correspond to an
interaction strength $U=-4V$, the dashed ones with
open symbols to $U=-2V$ and the dot-dashed ones to $U=0V$
($+,\times,*$).
Different symbols correspond to different linear size of
the lattice $L=20$
$(\circ,+)$, $L=30$ $(\Box,\times)$, and $L=40$
$(\bigtriangleup,\ast)$.
}
\label{fig2}
\end{figure}

In order to study the ground state properties of our model
in a more quantitative way it is convenient to compute the inverse
participation ratio (IPR) $\xi$ defined as 
$\xi^{-1}=\langle\sum_{\mathbf n}f^2({\mathbf n})\rangle$ where
the brackets mark the averaging over $N_D$ disorder realizations
(typically $N_D=100$). Physically, $\xi$ gives the number of lattice
sites occupied by one particle in the TIP ground state. The dependence
of the IPR $\xi$ on the disorder strength $W$ is shown in Fig.2 for
different strength of interaction $U$ and different system sizes 
$L$. In the absence of interaction, for finite system sizes used in
our numerical simulations ($L\leq 40$), the ground state is delocalized
for disorder $W\leq 5V$ and it becomes localized for $W>5V$. On the 
contrary in the presence of interaction the TIP ground state becomes
localized for $W> W_c \approx 2V$ at $U=-4V$ and for 
$W> W_c \approx 3V$ at $U=-2V$. Indeed for $W < W_c$ the IPR $\xi$
starts to grow significantly with the increase of the system size
$L$ that corresponds to pair delocalization.
The decrease of the $W_c$ value induced by the attractive interaction 
shows that the attraction leads to localization of pairs inside the 
non-interacting delocalized phase. This effect is absent in the mean
field approximation where the ground state remains well delocalized
(compare Figs. 1e and 1g). This phenomenon is similar to the situation
in three-dimensional Anderson model for which the localization
of pairs was discussed in \cite{lages}. As in \cite{lages} we
attribute this phenomenon to the increase of the effective mass 
$m_{eff}$ of the pair ($V \propto 1/m_{eff}$) that leads to a decrease
of the critical disorder strength ($W_c \sim V \propto 1/m_{eff}$).
For strong attraction the mass is approximately doubled so that the
value of $W_c$ is decreased by a factor of two comparing to the
non-interacting case. The numerical data in two and three dimensions
presented here and in \cite{lages} are in satisfactory agreement
with this estimate.
 
\begin{figure}
\epsfxsize=8cm
\epsfysize=8cm
\epsffile{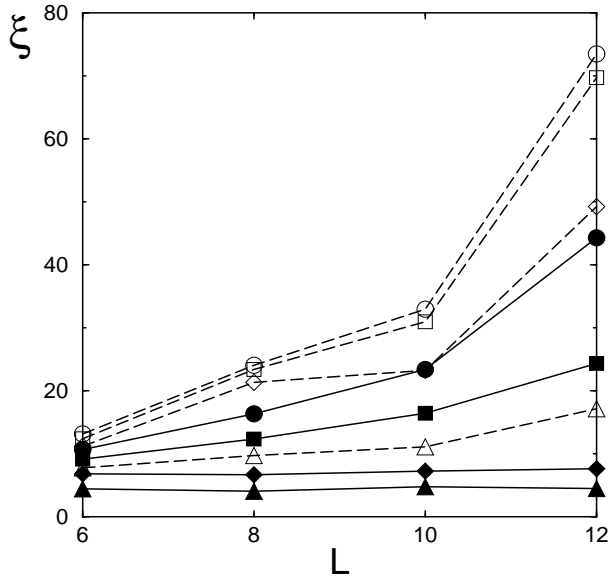}
\caption{IPR $\xi$ as a function of the linear size $L$.
Dashed lines and open symbols are for $W/V=2$ and
 full lines and full symbols are for $W/V=5V$,
with $U/V=0$ ($\circ$), $-1$ ($\Box$), $-2$ (diamond), $-4$
($\bigtriangleup$). The average is done
over 100 disorder realizations. Here, the filling factor is 
$\nu=1/4$ and $\alpha=L^2/M=2$.}
\label{fig3}
\end{figure}
The dependence of IPR $\xi$ on the system size $L$ for $W=2V$
and $W=5V$ is shown in Fig.3 for different values of interaction
$U$. For $L \leq 12$ the non-interacting states are delocalized.
The introduction of interaction decreases significantly the IPR
value and for $W=5V$ the TIP ground state is localized for 
$\arrowvert U\arrowvert \geq 2V$. On the contrary for $W=2V$
the IPR still grows with $L$ for $U=-2V$. This behavior is 
qualitatively similar to the numerical data obtained in 
\cite{bhargavi} by projected quantum Monte Carlo method 
(see Fig.3 there). According to \cite{bhargavi} the pairs at 
quarter filling become localized at $U/V\approx-4$ for $W=5V$
and remain delocalized for $W=2V$ (only sizes $L \leq 12$ where
accessible by this method). While the qualitative behavior is 
similar (compare Fig.3 here with Fig.3 in \cite{bhargavi})
the quantitative difference between the two sets of data is 
definitely present. For example in our Fig.3 at $W=5V$ the states 
become localized approximately at $U/V \approx -1.5$ and not 
at $U/V \approx -4$ as in \cite{bhargavi}. We attribute this 
quantitative differences to the fact that in \cite{bhargavi}
up to $74$ real spin fermions were present and were treated 
exactly (up to statistical errors) by the quantum Monte Carlo 
method. The presence of other fermions can renormalize effective
strength of interaction between two particles. Also it can 
change the effective strength of disorder for fermions near the 
Fermi level. The comparison between the two figures shows that 
the TIP approach captures the qualitative physical properties
of the system but quantitatively it gives different values.
In a sense this situation is similar to the comparison between
the Cooper approximation and the BCS theory.

\begin{figure}
\epsfxsize=8cm
\epsfysize=8cm
\epsffile{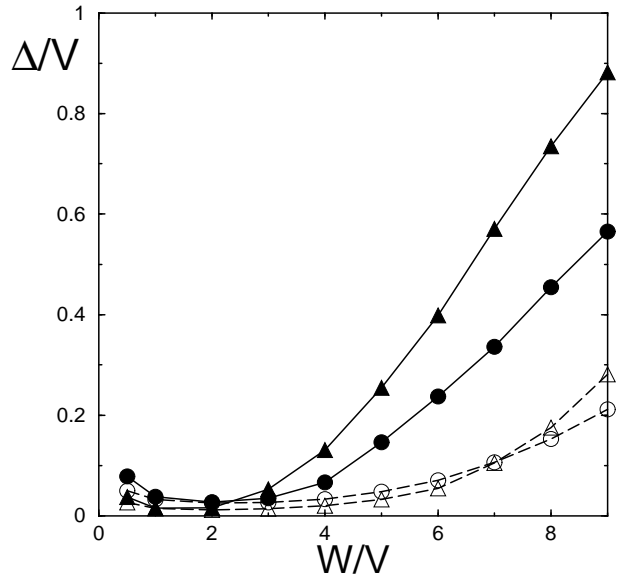}
\caption{Dependence of the coupling energy $\Delta$ on the disorder
strength $W$ for $U=-2V$. 
The linear lattice size is $L=20$ ($\circ$) and $L=40$ 
($\bigtriangleup$); data are obtained by exact diagonalization of TIP model 
(solid lines, full symbols) and by mean field approximation from 
the Cooper ansatz (dashed lines, empty symbols).}
\label{fig4}
\end{figure}

The difference between exact diagonalization of the TIP 
Hamiltonian (\ref{TIPSCH}) and the mean field solution given
by the Cooper ansatz is also clearly seen in the coupling 
energy of pair $\Delta = E_g(U=0)-E_g(U)$. Here $E_g$ is the
TIP ground state energy in presence of interaction $U$.
For $U=0$ we have $E_g(U=0)=2E_F$. The dependence of $\Delta$
on the disorder strength $W$ is shown in Fig.4 for $U=-2V$.
In the BLS phase at $W>W_c\approx 3V$ the coupling energy 
$\Delta$ obtained from exact diagonalization becomes 
significantly larger than the value of $\Delta$ given by 
the mean field approximation based on delocalized states.
This shows that energetically the BLS phase is more favorable
than the mean field Bogolubov-de Gennes solution \cite{BdG}.
The physical reason for the increase of $\Delta$ comparing 
to the mean field value is related to localization: the pairs
are localized and particles remains closer to each other
that effectively increases the coupling strength between them
\cite{note1}. On the contrary for $W<W_c$ when the pairs are 
delocalized the exact solution gives the values of $\Delta$
which are close to the mean field value. This is in agreement
with the Anderson theorem according to which mean field remains
valid in the regime with weak disorder.

The fact that for TIP the BLS phase is energetically more favorable 
than the mean field solution indicates that also at finite particle 
density the BLS phase will be more favorable. This indication
is in agreement with the quantum Monte Carlo computations presented
in \cite{bhargavi}. It is also possible to give another argument in 
the favor of BLS phase. Suppose that the filling factor $\nu$ is 
close to the critical value $\nu_c$ at the mobility edge of 
non-interacting particles (but $\nu>\nu_c$). Then it is natural that 
all particles below the mobility edge are localized. Then the
density of interacting pairs above $\nu_c$ is proportional to 
$\arrowvert \nu-\nu_c \arrowvert$ and is relatively low for
$\arrowvert \nu-\nu_c \arrowvert \ll \nu$. In this regime the 
pairs above $\nu_c$ are well separated and the TIP approximation 
we discuss in this paper should be rather reasonable. Of course
on the next step the residual interaction between pairs should
be taken into account \cite{note2}. In this picture it is clear that the BLS
phase is energetically more favorable comparing to the delocalized mean
field solution.

\section{Ground state properties with magnetic field}

It is interesting to understand how the TIP properties in the BLS phase
are affected by a magnetic field $\mathbf B$ perpendicular to the 
two-dimensional lattice. In this case the one-particle Hamiltonian 
takes the form
\begin{equation}\label{HAM}
H_{1}= \sum_{\mathbf n} E_{\mathbf n} \left\arrowvert
{\mathbf n}\left\rangle\right\langle{\mathbf n}\right\arrowvert
+V\sum_{\mathbf n}\left(T({\mathbf n})+
T^*({\mathbf n})\right)
\end{equation}
where  $T({\mathbf n})$ and $T^*({\mathbf n})$
are the translation operators from site ${\mathbf n}$
to  its nearest neighbours 
\begin{equation}
T({\mathbf n})=T_{M_{\mathbf x}}({\mathbf n})
\arrowvert{\mathbf n}\rangle\langle
{\mathbf n+e_x}\arrowvert+
T_{M_{\mathbf y}}({\mathbf n})\arrowvert{\mathbf n}\rangle
\langle{\mathbf n+e_y}\arrowvert.
\end{equation}
Here ${\mathbf e_x}$ and ${\mathbf e_y}$ are the unitary vectors
on the two dimensional lattice and 
\begin{equation}
T_{M_{\mathbf x(y)}}
({\mathbf n})=\exp\left(-\frac{iq}{\hbar c}
\int_{\Gamma_{\mathbf x(y)}({\mathbf n})}{\mathbf A}
.d{\mathbf n'}\right)
\end{equation}
are the magnetic translation operators along paths
$\Gamma_{\mathbf x}({\mathbf n})=({\mathbf n\rightarrow n+e_x})$ 
and $\Gamma_{\mathbf y}({\mathbf n})=({\mathbf n\rightarrow n+e_y})$.
For convenience
we choose the Landau's gauge for the magnetic field 
${\mathbf A}=-n_yB{\mathbf e_x}$. The magnetic translation
operators are then determined as $T_{M_{\mathbf x}}=
\exp(2\pi i\gamma n_y)$ and $T_{M_{\mathbf y}}=1$
with $\gamma=\frac{qB}{hc}$. Due to the periodic boundary
conditions the effective topology of the two-dimensional
lattice is that of a torus with a transversal and a longitudinal
radius $R_t=R_l=L$. This  topology implies flux quantization 
on the lattice 
so that $\gamma=\frac{m}{L}$ with $m\in [0,L-1]$. Then the one-particle
Hamiltonian (\ref{HAM}) can be explicitly written as
\begin{eqnarray}\label{HAM2}
H_{1}&&=\sum_{\mathbf n}E_{\mathbf n}\arrowvert{\mathbf n}
\rangle\langle{\mathbf n}\arrowvert \nonumber\\
&&+\sum_{\mathbf n}\Big(e^{2\pi i\gamma n_y}
\arrowvert{\mathbf n}\rangle\langle{\mathbf n+e_x}\arrowvert+
\arrowvert{\mathbf n}\rangle\langle{\mathbf n+e_y}\arrowvert
\Big)\nonumber\\
&& +\sum_{\mathbf n}\Big(e^{-2\pi i\gamma n_y}
\arrowvert{\mathbf n+e_x}\rangle
\langle{\mathbf n}\arrowvert+
\arrowvert{\mathbf n+e_y}\rangle
\langle{\mathbf n}\arrowvert\Big).
\end{eqnarray}

\begin{figure}
\epsfxsize=8cm
\epsfysize=11cm
\epsffile{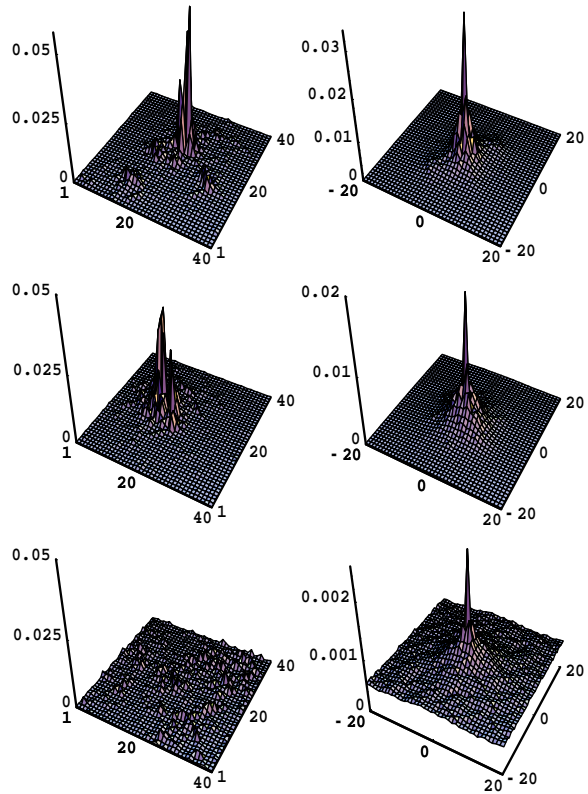}
\caption{Ground state probability distributions
of two interacting particles in a lattice of linear
size $L=40$, with disorder strength $W=3V$, and interaction 
strength $U=-2V$;
the left column shows the probability distribution
$f(\mathbf n)$ and the right one the interparticle
probability distribution $f_d(\mathbf r)$. The different
rows presents the TIP ground state for different value
of the magnetic flux $\gamma$:
top row $\gamma=0$, middle row $\gamma=\frac{2}{40}$,
and bottom row $\gamma=\frac{5}{40}$.
}
\label{fig5}
\end{figure}

We study now the ground state of the TIP Hamiltonian 
constructed with this new one-particle Hamiltonian $H_{1}$ (\ref{HAM2}).
This Hamiltonian is written in  the preferential
basis of  non-interacting  eigenstates of (\ref{HAM2})
that leads to the Schr\"odinger equation of the form (\ref{TIPSCH}).
As in the case with ${\mathbf B}={\mathbf 0}$ we use the probability 
distributions
$f({\mathbf n})$, $f_d({\mathbf r})$ and the IPR $\xi$ to
depict the ground state properties of the TIP problem in presence of a magnetic
field. Thus Fig.\ref{fig5} represents the  TIP probability distributions for a
system of linear size $L=40$ and fixed disorder and interaction
strengths ($W=3V,\, U=-2V$). The data are shown for
different magnetic flux ratios $\gamma=0,\,\gamma=\frac{2}{40}$ 
and $\gamma=\frac{5}{40}$. At $\gamma=0$
the one particle probability is well localized by interaction
(first row of Fig.\ref{fig5}). However, with the increase of magnetic flux
$\gamma$ the localization is destroyed. The data of Fig.\ref{fig5}
suggest that there exists a critical magnetic flux
$\gamma_c$ below which the TIP pairs remain localized ($\gamma<\gamma_c$,
middle row of Fig.\ref{fig5}) and above which pairs become totally delocalized
($\gamma>\gamma_c$, bottom row of Fig.\ref{fig5}). At the same time 
for $\gamma>\gamma_c$ the interparticle distance probability
distribution $f_d({\mathbf r})$ is less peaked. Hence for $\gamma>\gamma_c$
the size of the pair is significantly increased.

\begin{figure}
\epsfxsize=8cm
\epsfysize=8cm
\epsffile{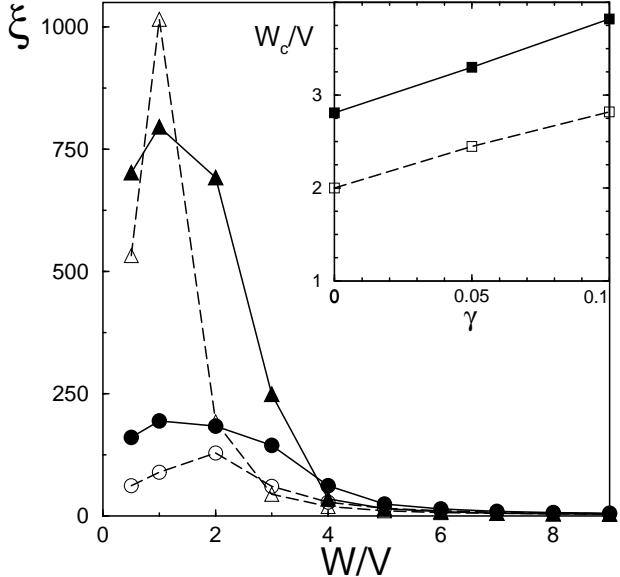}
\caption{Dependence of the inverse participating ratio 
$\xi$ of the TIP ground state on the disorder $W/V$ and
magnetic flux $\gamma$. The interaction 
strength 
for the main figure is $U=-2V$. Empty/full symbols
correspond to $\gamma=0/\gamma=0.1$, with different
linear lattice size $L=20$ ($\circ$)
and $L=40$ ($\bigtriangleup$).
Inset shows the  phase diagram in the plane of $W_c/V$ and $\gamma$; 
the full line corresponds to $U=-2V$ and the dashed line to $U=-4V$;
BLS phase is at $W>W_c$.
}
\label{fig6}
\end{figure}

The ground state properties can be studied in a more
quantitative way with the help of the IPR $\xi$ defined above.
Fig.\ref{fig6} represents the dependence of $\xi$ on the disorder
strength $W$ for a fixed interaction $U=-2V$. Data are shown
for different values of magnetic flux $\gamma$. They clearly show
that the introduction of  magnetic field leads to an  increase
of $\xi$ at a fixed value of $W$. Thus the magnetic field
enhances the delocalization of particles for $\gamma > \gamma_c(U)$
and $W<W_c(U)$. On the contrary for $\gamma < \gamma_c(U)$
and $W > W_c(U)$ the variation of $\xi$ with lattice size $L$
is weak and usually here there is a small
decrease of $\xi$ with increase of $L$ (see Fig.\ref{fig6}).
The delocalization transition can be determined as the
point where $\xi$ is independent of the lattice size
(crossing point). An approximate phase diagram in the plane $(W,\gamma)$
obtained in this way is shown in the inset of Fig.\ref{fig6}.
With the increase of 
\begin{figure}
\epsfxsize=8cm
\epsfysize=8cm
\epsffile{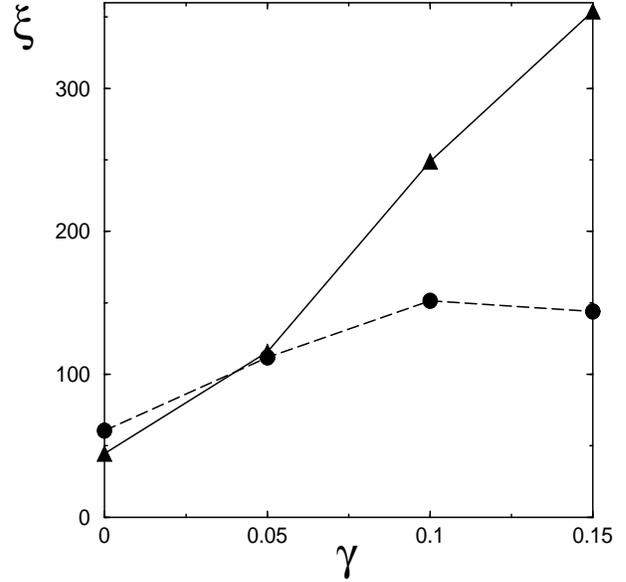}
\caption{Dependence of the inverse participation ratio 
$\xi$ of the TIP ground state on the magnetic flux
$\gamma$ for two lattice sizes $L=20$ $(\bullet)$ and 
$L=40$ (triangle).
Here the interaction strength is $U=-2V$ 
and the disorder strength is $W=3V$.
}
\label{fig7}
\end{figure}

\begin{figure}
\epsfxsize=8cm
\epsfysize=8cm
\epsffile{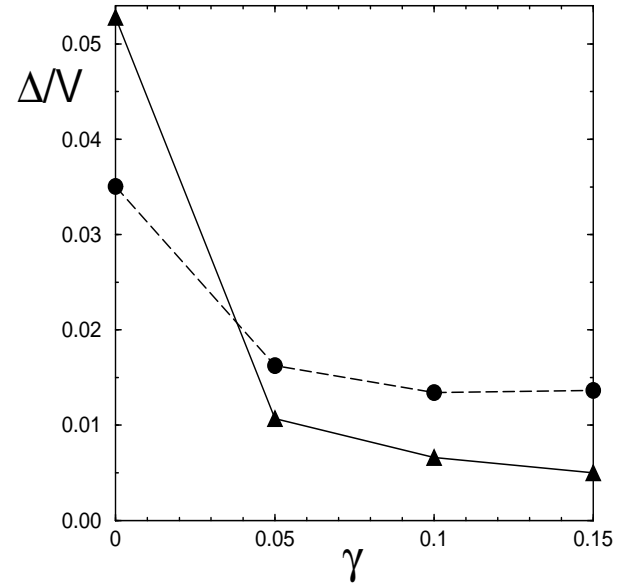}
\caption{Pair coupling energy $\Delta$ as a function of  magnetic flux
$\gamma$ for  $L=20$ $(\bullet)$
and $L=40$ (triangle). The interaction and disorder
strengths  are as in Fig.\ref{fig7}: $U=-2V$ and  $W=3V$.
}
\label{fig8}
\end{figure}
\noindent interaction the localized phase penetrates
deeper in the region of weak disorder.
The fact that a magnetic field can delocalize pairs
inside the BLS phase is also illustrated in Fig.\ref{fig7}
where the IPR $\xi$ is enormously increased
by the magnetic flux. This result is in agreement with a general
fact known for non-interacting particles that the localization length 
is increased by a magnetic field \cite{mirlin}. While the delocalization 
induced by a magnetic
field is clearly illustrated by the Figs.\ref{fig6} and \ref{fig7}
it is rather difficult to determine numerically the properties
of pairs in the delocalized phase. In this phase an effective
pair size becomes too large to investigate it numerically.
The question if the superconductivity survives or 
if a magnetic field drives the system to a metallic regime
is difficult to answer in the frame of our numerical 
approach.

For a better understanding of both localized and delocalized 
phases we studied the dependence of the pair coupling
energy $\Delta$ on the strength of magnetic field. 
In the standard Cooper problem $\Delta$
is related to the BCS gap and determines the Cooper
pair size $l_{pair}\propto 1/\Delta$. Fig.\ref{fig8}
shows the dependence of $\Delta$ on the magnetic field for
different lattice sizes and for the case $U=-2V$ and $W=3V$, already
used in the Fig.\ref{fig7}. In the localized regime with
$\gamma < \gamma_c \approx 0.04$ the value of $\Delta$ 
varies weakly with the growth of $L$. In contrast, for
$\gamma > \gamma_c$ its value decreases in 2 - 3 times.
This can be considered as an indication that
the superconductivity is significantly suppressed by 
magnetic field for $\gamma > \gamma_c$.
However, a significant increase of $L$ is required to
investigate the properties of pairs in the delocalized
regime.

\section{Conclusion}

The present studies show that in the presence of disorder
the attractive Hubbard interaction leads to localization
of pairs and appearance of phase with bi-particle localized states
which is located inside the non-interacting metallic regime.
This BLS phase cannot be obtained in the mean-field 
approximation.
It is shown that it can be destroyed by the introduction
of a magnetic field which drives the system to delocalization.
In the BLS phase the pair coupling energy is much
larger than the value obtained in the mean-field 
approximation. This indicates that  the BLS phase is 
energetically more preferable comparing to the mean-field
solution. The results obtained for two particles (one pair)
are in qualitative agreement with the recent
results obtained with the quantum Monte Carlo in \cite{bhargavi}.

It is our pleasure to thank G.Benenti and B.Srinivasan for
stimulating discussions.
We thank the IDRIS in Orsay and the CalMiP in Toulouse for access to 
their supercomputers.


\begin{thebibliography}{99}
\bibitem[*]{byline1} http://w3-phystheo.ups-tlse.fr/$\sim$dima
\bibitem{larkin}     A. Larkin, Ann. Phys.(Leipzig) {\bf 8}, 785 (1999).
\bibitem{matveev}    K.A. Matveev and A.I. Larkin, Phys Rev. Lett. 
                     {\bf 78}, 3749 (1997).
\bibitem{erik94}     M. Wallin, E. S. S\o rensen, S. M. Girvin, 
                     and A. P. Young, Phys. Rev. B {\bf 49}, 
                     12115 (1994).
\bibitem{trivedi1}   R.T. Scalettar, N. Trivedi and C. Huscroft,
                     Phys. Rev. B {\bf 59}, 4364 (1999).
\bibitem{franz}      M. Franz, C. Kallin, A.J. Berlinsky and
                     M.I. Salkola, Phys. Rev. B {\bf 56}, 7882 
                     (1997).
\bibitem{trivedi2}   A. Ghosal, M. Randeria, and N. Trivedi,
                     Phys. Rev. B {\bf 63}, 020505(R) (2001).
\bibitem{trivedi3}   P.J.H. Denteneer, R.T. Scalettar, and N. Trivedi,
                     Phys. Rev. Lett. {\bf 83}, 4610 (1999).
\bibitem{goldman}    A.M. Goldman and N. Markovi\'c, Physics Today
                     {\bf 51}, 39 (1998).
\bibitem{gantmakher} V.F. Gantmakher, M.V. Golubkov, V.T. Dolgopolov,
                     G.E. Tsydynzhapov, and A.A. Shashkin,
                     JETP Lett. {\bf 68}, 363 (1998).
\bibitem{boebinger}  G.S. Boebinger, Y. Ando, A. Passner,
                     T. Kimura, M. Okuya, J. Shimoyama, K. Kishio,
                     K. Tamasaku, N. Ichikawa, and S. Uchida,
                     Phys. Rev. Lett. {\bf 77}, 5417 (1996).
\bibitem{ando}       K. Segawa and Y. Ando, J. Low Temp. Phys. {\bf 117}, 
                     1175 (1999)
\bibitem{anderson}   P.W. Anderson, J. Phys. Chem. Solid {\bf 11},
                     26 (1959).
\bibitem{abrikosov}  A.A. Abrikosov and L.P. Gorkov, Sov. Phys. 
                     JETP {\bf 8}, 1090 (1958).
\bibitem{poilblanc}  G. Montambaux, D. Poilblanc, J. Bellissard, 
                     and C. Sire, Phys. Rev. Lett. {\bf 70}, 497 (1993).
\bibitem{dagota}     E. Dagotto, Rev. Mod. Phys. {\bf 66}, 763 (1994).
\bibitem{cooper}     L.N. Cooper,  Phys. Rev. {\bf 104}, 1189 (1956).
\bibitem{lages}      J. Lages and D.L. Shepelyansky, Phys. Rev. B
                     {\bf 62}, 8665 (2000).
\bibitem{bhargavi}   B. Srinivasan and D.L. Shepelyansky, cond-mat/0102055.
\bibitem{mirlin}     A. Mirlin, Phys. Rep. {\bf 326}, 259 (2000).
\bibitem{imry}       Y. Imry, Europhys. Lett. {\bf 30}, 405 (1995).
\bibitem{jacquod}    Ph. Jacquod and D.L. Shepelyansky, Phys. Rev. Lett. 
                     {\bf 78}, 4986 (1997).
\bibitem{oppen}      F. von Oppen, T. Wettig, and J. M\"uller,
                     Phys. Rev. Lett. {\bf 76}, 491 (1996);
                     and p.235 in {\it Correlated fermions and
                     transport in mesoscopic systems}, Eds. T.Martin, 
                     G.Montambaux and J.Tr\^an Thanh V\^an, Editions 
                     Frontieres, Gif-sur-Yvette, 1 - 553 (1996).
\bibitem{1dtip}      M.A.N. Ara\'ujo, M. Dzierzawa, C. Aebischer, 
                     D. Baeriswyl,
                     Physica B {\bf 244}, 9 (1998).
\bibitem{BdG}        P.G. de Gennes, {\it Superconductivity of Metals and
                     Alloys}, W.A. Benjamin, Inc., New York (1966).
\bibitem{note1}      In the BLS phase at $W>3V$ the value of $\Delta$ 
                     obtained by exact diagonalization (full symbols)
                     demonstrates some increase with the change of $L$
                     from $L=20$ to $L=40$. We attribute this to the
                     fact that at larger $L$ the pair in the ground state 
                     can choose a more profound minimum appearing due to 
                     statistical fluctuations that gives some growth of 
                     $\Delta$.
\bibitem{note2}      The first step in direct numerical investigations of 
                     few interacting spin fermions above the frozen Fermi 
                     sea was done in J. Lages, G. Benenti, and 
                     D.L. Shepelyansky, cond-mat/0101265. However in this
                     approach it is difficult to reach large system sizes
                     since the condition $M/L^2=const$ leads to a strong 
                     growth of the matrix size with the number of particles.
\end{thebibliography}
\end{document}